\documentclass[aps,pra,twocolumn,superscriptaddress]{revtex4-1}

\usepackage{graphicx}
\usepackage{hyperref}
\usepackage{amsmath}
\usepackage{amssymb}
\usepackage{epstopdf}
\usepackage{multirow}
\usepackage{threeparttable}
\usepackage[usenames]{color}
\usepackage{xcolor}
\usepackage[normalem]{ulem} 

\begin{document}
	
\title{Mid-infrared temporal ghost imaging via two-photon structured encoding}

\author{Ziyu He}
\affiliation{State Key Laboratory of Precision Spectroscopy, and Hainan Institute, East China Normal University, Shanghai 200062, China}
	
\author{Kun Huang}
\email{khuang@lps.ecnu.edu.cn}
\affiliation{State Key Laboratory of Precision Spectroscopy, and Hainan Institute, East China Normal University, Shanghai 200062, China}
\affiliation{Chongqing Key Laboratory of Precision Optics, Chongqing Institute of East China Normal University, Chongqing 401121, China}
\affiliation{Collaborative Innovation Center of Extreme Optics, Shanxi University, Taiyuan, Shanxi 030006, China}

\author{Huijie Ma}
\affiliation{State Key Laboratory of Precision Spectroscopy, and Hainan Institute, East China Normal University, Shanghai 200062, China}

\author{Wen Zhang}
\affiliation{State Key Laboratory of Precision Spectroscopy, and Hainan Institute, East China Normal University, Shanghai 200062, China}

\author{Jianan Fang}
\affiliation{State Key Laboratory of Precision Spectroscopy, and Hainan Institute, East China Normal University, Shanghai 200062, China}
\affiliation{Chongqing Key Laboratory of Precision Optics, Chongqing Institute of East China Normal University, Chongqing 401121, China}

\author{Heping Zeng}
\email{hpzeng@phy.ecnu.edu.cn}
\affiliation{State Key Laboratory of Precision Spectroscopy, and Hainan Institute, East China Normal University, Shanghai 200062, China}
\affiliation{Chongqing Key Laboratory of Precision Optics, Chongqing Institute of East China Normal University, Chongqing 401121, China}
\affiliation{Shanghai Research Center for Quantum Sciences, Shanghai 201315, China}
\affiliation{Chongqing Institute for Brain and Intelligence, Guangyang Bay Laboratory, Chongqing, 400064, China}

\begin{abstract}	
Temporal ghost imaging (TGI) enables ultrafast signal reconstruction beyond electronic bandwidth limits. Extending this concept to the mid-infrared (MIR) regime through nonlinear frequency conversion offers new opportunities for high-fidelity temporal detection, but remains constrained by stringent phase-matching condition, limited spectral coverage, and intricate optical alignment. Here, we propose and demonstrate a broadband MIR TGI system based on non-degenerate two-photon absorption. A temporally encoded near-infrared pump transfers structured modulation onto a MIR signal directly at a silicon detector, which facilitates concurrent modulation and detection without external nonlinear crystals. The reconstructed temporal waveforms exceed the detector bandwidth by more than fortyfold, achieve a detection sensitivity of 0.05 pJ/pulse, allow compressed sensing with 80\% fewer measurements, and support broadband operation across 2.5-3.8 $\mu$m. This compact, alignment-free, and room-temperature system establishes a practical route for fast and sensitive MIR time-domain analysis, holding great promise for applications in time-resolved molecular spectroscopy, high-precision infrared ranging, and high-speed free-space communication.
\end{abstract}

\maketitle
	
\section{Introduction}
Temporal ghost imaging (TGI) is an emerging technique that reconstructs ultrafast temporal signals through intensity correlations without requiring direct time-resolved detection \cite{Faccio2016NP}. From the perspective of space-time duality \cite{Salem2013AOP}, TGI can be viewed as the temporal analogue of spatial ghost imaging, in which information is retrieved through correlations between structured illumination and integrated detection \cite{Erkmen2010AOP}. This correspondence endows TGI with the inherent features of ghost imaging, including robustness against noise, reduced susceptibility to distortions, and enhanced information security \cite{Gatti2004PRL}. By correlating random or structured temporal sequences with the integrated outputs of a slow ``bucket'' detector, TGI effectively overcomes the electronic bandwidth limitation of conventional photodetectors, enabling access to ultrafast temporal structures otherwise undetectable \cite{Ryczkowski2016NP}. Owing to these advantages, TGI has found applications in long-distance communication \cite{Yao2018OL, Chen2021OL}, secure data transmission \cite{Jiang2017SR}, and quantum information processing \cite{Wu2019OL}.

Early implementations of TGI relied on random fluctuations from chaotic lasers \cite{Ryczkowski2016NP, Wu2020OE}, spontaneous parametric downconversion (SPDC) photon pairs \cite{Denis2017JO}, or thermal light sources \cite{Devaux2016JO, Yao2018OL}, typically requiring a reference arm for correlation and thus suffering from limited efficiency and stability. The advent of computational TGI introduced programmable light modulators to generate deterministic encoding sequences, eliminating the reference arm and improving reproducibility \cite{Xu2018OE}. The adoption of orthogonal encoding bases such as Walsh-Hadamard matrices enabled one to improve reconstruction fidelity, computational efficiency, and noise resilience in comparison with random encoding \cite{Devaux2016Optica}. Furthermore, enhanced variants including differential \cite{Oka2017APL} and amplified TGI \cite{Ryczkowski2017APLP} further boosted signal-to-noise ratio and temporal resolution, while Fourier TGI \cite{Wenwen2020OLE} provides complementary spectral information. Nevertheless, most demonstrations have remained in the visible and near-infrared (NIR) regimes, where mature modulators and detectors are readily available.

Indeed, extending TGI into other spectral regions has been hampered by the lack of high-speed modulators and efficient detectors, particularly in the mid-infrared (MIR, 2-20 $\mu$m) range. This regime is of great importance, as it encompasses the fundamental vibrational fingerprints of molecules, providing rich spectroscopic information for chemical sensing, environmental monitoring, and biomedical diagnostics \cite{Cheng2015Science, Shi2020NM, Razeghi2014RPP}. To overcome detection challenges, various indirect MIR detection schemes have been explored based on nonlinear wavelength conversion, including optical parametric upconversion \cite{Dam2012NP, Huang2022NC, Junaid2019Optica, Barh2019AOP} and quantum correlation-based methods \cite{Paterova2020SA, Kviatkovsky2020SA}. These approaches have enabled remarkable progress in the spatial domain for MIR single-photon imaging \cite{Huang2022NC} and hyperspectral imaging \cite{Fang2024NC}, as well as in the spectral domain for MIR spectroscopy \cite{Liu2023NC, Cai2024SA}.  In the temporal domain, pioneering studies of MIR TGI employed two-color second-harmonic generation for upconverting MIR fluctuations into the NIR, achieving temporal imaging at 2 $\mu$m \cite{Wu2019Optica}. Subsequent demonstrations based on difference-frequency generation (DFG) transferred temporal modulation from the NIR pump to the MIR signal, overcoming the lack of high-speed MIR modulators \cite{Wu2024LSA}. More recently, preprogrammed upconversion methods have further improved sensitivity and reconstruction fidelity, enabling MIR single-photon TGI at 3.4 $\mu$m \cite{Zhang2025LPR}. Despite great advances, the crystal-based nonlinear conversion scheme remains constrained by stringent phase-matching requirements, which impose limited spectral coverage and demand careful optical alignment. Furthermore, the involved modulation and detection are implemented in separate stages, fundamentally restricting compactness and scalable integration. 

Alternatively, an effective route for sensitive MIR detection can leverage nonlinear carrier dynamics in semiconductors, such as two-photon absorption (TPA) \cite{Cirloganu2011OE, Fishman2011NP} and higher-order multiphoton processes \cite{Nevet2011OL, Pearl2008APL}. In particular, non-degenerate two-photon absorption (ND-TPA) provides enhanced nonlinear coefficient, and inherently supports broadband MIR operation, enabling modulation transfer and detection without requiring phase matching \cite{Sterczewski2023NC, Fishman2011NP}. ND-TPA based detection has been successfully applied to room-temperature MIR sensing \cite{Fang2020PRA, Fix2017APL}, video-rate spectral imaging \cite{Knez2020LSA, Knez2022SA, Potma2021APLP}, high-sensitivity single-pixel imaging \cite{Ma2025PhotoniX}, and high-precision depth-resolved tomography \cite{Potma2021Optica, Potma2021APLP}. These advances demonstrate the feasibility of ND-TPA for broadband, compact, and room-temperature MIR detection, underscoring its strong potential as a foundation for extending TGI into the MIR regime.

Here, we demonstrate a broadband MIR TGI system enabled by ND-TPA in a silicon photodiode. A 1550 nm pump carrying preprogrammed temporal patterns transfers structured modulation onto the MIR signal directly at the detector. Through the ND-TPA process, the MIR temporal modulation and upconversion detection occur simultaneously within the same device, eliminating the need for external nonlinear crystals and complex optical alignment. This concurrent modulation-detection operation intrinsically enhances system compactness, stability, and scalability, establishing a unified platform for efficient MIR temporal measurement. The scheme achieves temporal reconstructions exceeding the detector bandwidth by more than fortyfold, supports sub-Nyquist compressed sensing with 80\% fewer measurements, and reaches a detection sensitivity of 0.05 pJ per pulse. Leveraging the broadband nature of ND-TPA, the system operates across 2.5-3.8 $\mu$m without the need to adjust system parameters. These results underscore the distinctive advantage of concurrent modulation-detection operation, positioning ND-TPA-enabled MIR TGI as a viable and scalable platform for fast and sensitive MIR temporal analysis over a wide spectral coverage.

\section{Basic principle}
\begin{figure*}[t!]
\includegraphics[width=0.9\textwidth]{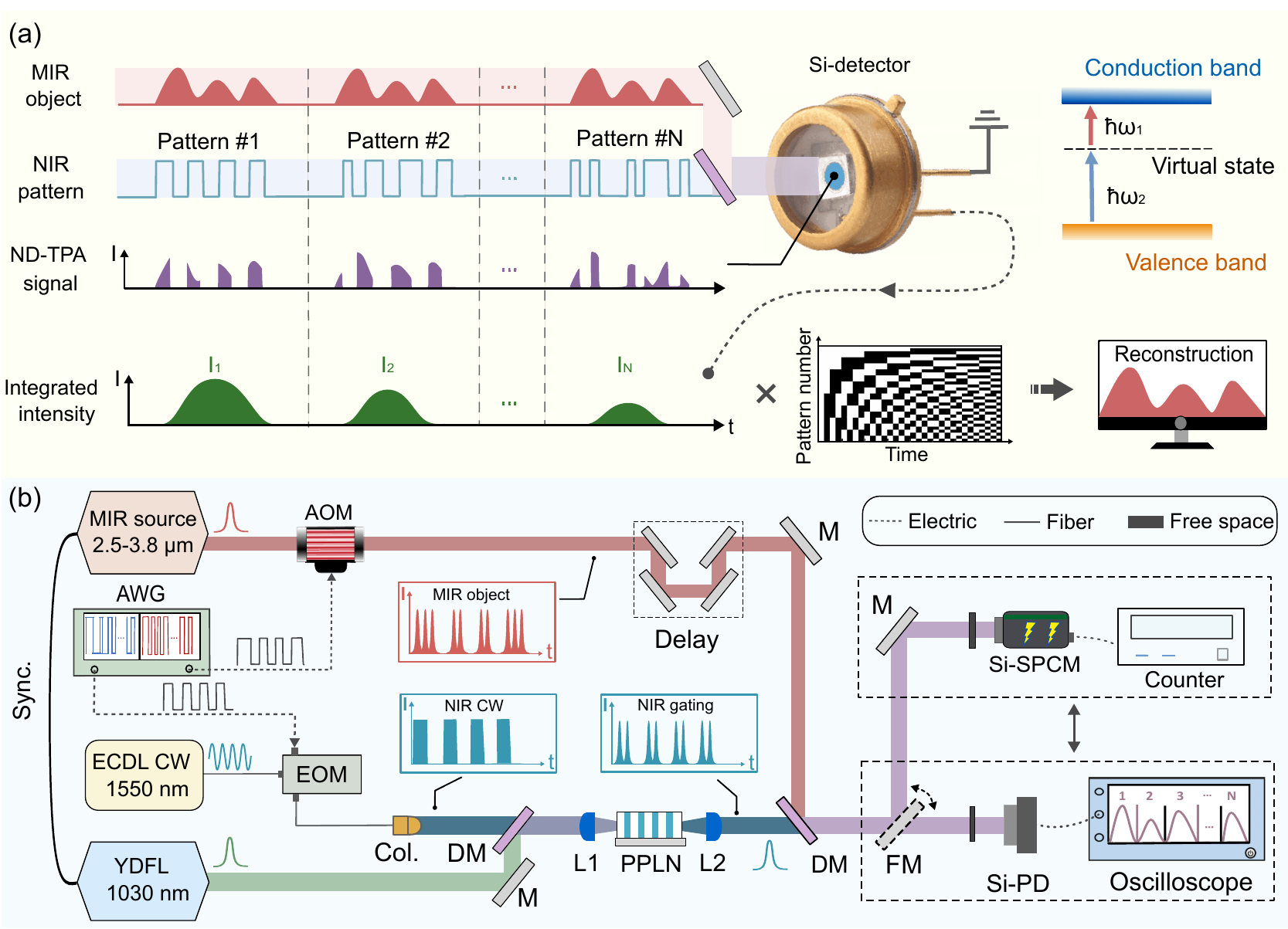}
\caption{Concept and implementation of the broadband MIR TGI system based on ND-TPA structured encoding. (a) Conceptual illustration of the operating principle. The near-infrared (NIR) pump is temporally modulated according to a Walsh-Hadamard matrix, while the mid-infrared (MIR) signal is intensity-encoded to emulate a temporal object. At a silicon detector, the combined NIR and MIR beams interact via the ND-TPA process, generating an electrical response as depicted in the energy-level diagram. The MIR temporal object is then reconstructed by correlating the integrated detector voltages with the predefined structured modulation patterns. (b) Schematic of the experimental setup. A broadly tunable MIR beam passes through a temporal object emulated by an acousto-optic modulator (AOM), whose transmission profile is controlled by an arbitrary waveform generator (AWG). A continuous-wave external-cavity diode laser (ECDL) at 1550 nm is temporally modulated by an electro-optic modulator (EOM) driven by the AWG, and subsequently amplified via optical parametric amplification in a periodically poled lithium niobate (PPLN) crystal pumped at 1030 nm. The generated NIR pulse train is temporally synchronized to the MIR signal, and the gating patterns are transferred to the intensity envelopes of the pulsed pump. The MIR and NIR beams are combined to focus onto a silicon photodiode (Si-PD) or single-photon counting module (SPCM), where the ND-TPA signals are recorded by a digital oscilloscope or a  frequency counter, respectively. Correlating the measured integrated intensities with the preprogrammed patterns enables recovery of the MIR temporal waveform. Abbreviations: M, silver mirror; DM, dichroic mirror; Col., fiber collimator; FM, flipping mirror.}
\label{fig1}
\end{figure*}

Figure \ref{fig1}(a) illustrates the operating principle of the broadband MIR TGI system based on ND-TPA. The system consists of three main parts: pump light modulation, ND-TPA detection, and temporal object reconstruction.
The light-source modulation module provides the temporal encoding basis. By leveraging high-performance NIR modulators, preprogrammed temporal patterns are imposed on the NIR pump beam. These patterns interact with the MIR waveform via ND-TPA at the detector surface, acting as a temporal gate and producing an ND-TPA signal that modulates the temporal characteristics of the MIR waveform. This indirect modulation strategy enables high-speed, flexible, and all-optical encoding of MIR temporal information.

Specifically, ND-TPA detection forms the core of the system. The TPA process can occur in two regimes: degenerate TPA, where both photons have identical energies ($\omega_1 = \omega_2$), and ND-TPA, where the photons possess different energies ($\omega_1 \neq \omega_2$). ND-TPA typically exhibits a higher nonlinear absorption coefficient due to resonant enhancement, where one photon is closer in energy to a real electronic transition, thereby increasing the effective transition probability \cite{Fishman2011NP}.

In our implementation, the NIR pump and MIR signal are spatially and temporally overlapped at a silicon detector. ND-TPA generates carriers within the detector material, and the resulting photocurrent is recorded. The underlying mechanism can be described as follows: one photon excites an electron from the valence band to a short-lived virtual state, where the electron can remain only for a duration limited by the Heisenberg uncertainty relation $\Delta t = \hbar / \Delta E$, with $\Delta E$ denoting the detuning from a real electronic level \cite{Fishman2011NP}. If, within this femtosecond-scale window, a second photon is absorbed such that the energy conservation condition is satisfied, 
\begin{equation}
\hbar \omega_s + \hbar \omega_p \geq E_g \ ,
\label{eq1}
\end{equation}
where $E_g$ is the semiconductor bandgap, the electron is promoted into the conduction band. In ND-TPA, the detected signal intensity scales with the product of the instantaneous intensities of the two incident beams,
\begin{equation}
I_{\mathrm{ND\text{-}TPA}}(t) \propto I_s(t) I_p(t) \ ,
\label{eq2}
\end{equation}
where $I_s$ and $I_p$ denote the intensities of the MIR signal and NIR pump, respectively. In the experiments, the detector operates in the unsaturated ND-TPA regime and maintains stable spatiotemporal overlap between the NIR pump and the MIR signal, so that the ND-TPA signal is proportional to the product of the NIR-MIR intensities. This nonlinear interaction allows the temporal modulation of the NIR pump to be faithfully transferred to the MIR signal through the ND-TPA process. In this configuration, the structured NIR waveform deterministically encodes the MIR signal directly at the detector.

Finally, the temporal object is retrieved by correlating the detector outputs with the preprogrammed encoding patterns. The Walsh-Hadamard matrix is used as the encoding patterns, which provides an efficient and noise-robust bases for TGI. The temporal object is discretized into $N$ sampling points and represented by a column vector $\mathbf{O} \in \mathbb{R}^{N \times 1}$. The encoding basis is an $N \times N$ Walsh-Hadamard matrix $\Phi \in \{-1, +1\}^{N \times N}$, with each row corresponding to one temporal pattern applied to the NIR pump. Because the Walsh-Hadamard matrix contains both $+1$ and $-1$ entries, it cannot be directly realized with optical modulators that produce non-negative intensities. Hence, each row of $\Phi$ is decomposed into two complementary non-negative sub-patterns, $\Phi^o$ and $\Phi^e$. A differential measurement is obtained via the intensity subtraction: $\mathbf{I} = (\Phi^o - \Phi^e)\mathbf{O}$. This differential scheme not only makes Hadamard modulation experimentally realizable but also cancels common-mode noise and preserves the orthogonality of the encoding basis, ensuring robust, high-fidelity reconstruction \cite{Devaux2016Optica, Ye2021PRA}. The temporal waveform is then recovered via the inverse transform,
\begin{equation}
\mathit{O} = (\Phi^o - \Phi^e)^{-1}\mathbf{I} \ .
\label{eq5}
\end{equation}

Moreover, by exploiting the sparsity of temporal signals, accurate waveform reconstruction can be achieved from fewer than $N$ measurements, enabling sub-Nyquist sampling and improving acquisition efficiency \cite{Zhang2025LPR}. This compressive strategy not only reduces the data acquisition time and hardware burden, but also enhances noise robustness by avoiding redundant measurements.

\begin{figure*}[t!]
\includegraphics[width=0.9\textwidth]{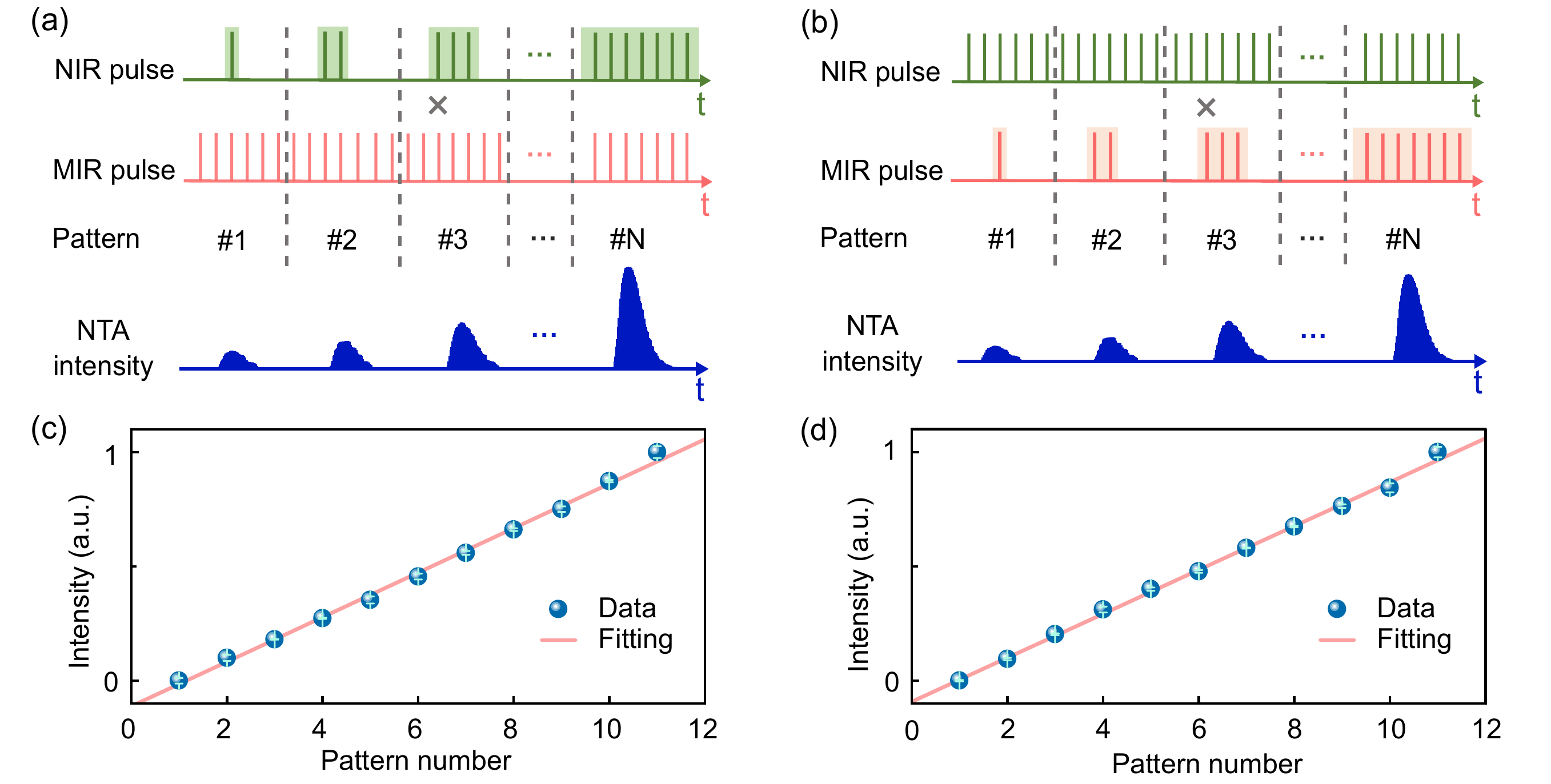}
\caption{Characterization of the ND-TPA detector response under different modulation conditions. (a) Pump modulation: The NIR gating pulses are temporally modulated, while the MIR signal pulses remain in the all-on state. (b) Signal modulation: The MIR pulses are temporally modulated, while the NIR pulses are held constant. (c) Integrated ND-TPA intensities under NIR modulation, showing linear growth with the increasing number of ``on'' bits.
(d) Integrated ND-TPA intensities under MIR modulation, likewise exhibiting consistent linear dependence. The observed bilinear response of the ND-TPA process to both pump and signal modulations lays the foundation for high-fidelity reconstruction in the TGI system. Note that error bars represent standard deviations from five consecutive measurements.}
\label{fig2}
\end{figure*}

\section{Experimental setup}
Figure \ref{fig1}(b) shows the schematic of the broadband MIR TGI system based on structured-pump ND-TPA. The system consists of two temporally synchronized beams: a modulation-encoded NIR pump and an intensity-encoded MIR signal. Both beams are derived from a common Yb-doped fiber laser (YDFL, LangyanTech YbFemto ProH) operating at 1030 nm, delivering 1 ps pulses at a repetition rate of 20.1 MHz. The laser output is divided into two branches for NIR pump generation and MIR signal generation, respectively.

In the NIR branch, one portion of the YDFL output is directed to an optical parametric amplifier (OPA) for pump generation. A continuous-wave (CW) seed at 1550 nm from an external-cavity diode laser is first temporally modulated by an electro-optic modulator (EOM), driven by preprogrammed Walsh-Hadamard sequences generated from an arbitrary waveform generator (AWG, RIGOL DG4162). The modulated seed and 1030 nm pump pulses are then co-injected into a periodically poled lithium niobate (PPLN) crystal. The seeded amplification suppresses spontaneous parametric downconversion, producing a modulation-encoded 1550 nm pulse train with an average output power of approximately 10 mW. This output serves as the structured NIR gating beam. By imprinting the temporal codes onto the fiber-coupled CW seed prior to amplification, additional free-space coupling is avoided and modulator- related insertion loss is miti gated in the overall power budget, helping preserve the encoded waveform during subsequent power scaling.

In the MIR branch, the other portion of the YDFL output pumps a fiber-feedback optical parametric oscillator (FOPO), generating stable and tunable MIR pulses spanning 2.5-3.8 $\mu$m with an output pulse width of 1 ps \cite{Yu2024PR}. After passing through a 2.4 $\mu$m long-pass filter, the MIR beam delivers an average power of 85 mW and is temporally modulated by an acousto-optic modulator (AOM), also driven by the AWG. The AOM modulation defines the temporal object used for TGI reconstruction.

Notably, passive synchronization between the two beams ensures coincident arrival of the MIR and NIR pulses at the detector. Fine temporal alignment is achieved by adjusting an optical delay line, while spatial overlap is optimized using dichroic mirrors to maximize ND-TPA efficiency. The nonlinear interaction occurs at a silicon photodiode (Si-APD, Thorlabs APD440A, 400-1100 nm, 100 kHz bandwidth), where the simultaneous absorption of an MIR and an NIR photon generates a measurable photocurrent. The integrated ND-TPA signals are recorded by a real-time oscilloscope (LeCroy WaveSurfer 3034Z).

For low-light and single-photon measurements, a silicon single-photon counting module (SPCM, Laser Components SAP500T6) connected to a frequency counter (Tektronix FCA3100) is used to register photon counts under different modulation patterns. A silicon window (LBTEK OW2-Si) in front of the detector eliminates ambient background noises. By correlating the sequence of recorded ND-TPA intensities with the preprogrammed modulation patterns, the temporal profile of the MIR object can be reconstructed.

\section{Results and discussion}
\subsection{Characterization of ND-TPA detector response}
Accurate TGI requires a detector with linear intensity response and sufficient temporal resolution to faithfully capture the encoded temporal patterns. In our MIR TGI system, the Walsh-Hadamard modulation was applied to the NIR pump, which then modulated the MIR signal through nonlinear interaction. The integrated ND-TPA intensities were subsequently correlated with the encoding matrix to reconstruct the temporal object. Since each temporal code corresponds to a distinct number of ``on'' bits, characterizing the linearity of detector is essential for reliable mapping between the encoded patterns and the measured ND-TPA signals.

We first evaluated the ND-TPA detector response under two complementary modulation configurations. As shown in Fig. \ref{fig2}(a), the MIR pulses were kept in an all-on state while the NIR pump was temporally modulated using an EOM driven by preprogrammed patterns. Conversely, in Fig. \ref{fig2}(b), the NIR pump remained constant while the MIR pulses were modulated by an AOM. Here, the interval between adjacent patterns was set to longer than 10 $\mu$s, which is determined by the 100 kHz photodiode readout bandwidth. This interval ensures pattern resolved settling of the detector output and avoids inter pattern carryover. This choice is not constrained by the ND-TPA response time, which is on the picosecond scale set by the pump-MIR temporal overlap. Figures \ref{fig2}(c-d) show the measured integrated ND-TPA intensities for each modulation pattern. Because the number of ``on'' bits in the modulation sequence increases linearly with pattern order, a proportional rise in ND-TPA signal is expected and experimentally confirmed. The observed bilinear behavior of the ND-TPA process with respect to both NIR and MIR modulations validates the detector's linearity, which formed the physical basis for high-fidelity temporal reconstruction in TGI.

\begin{figure}[t!]
\includegraphics[width=1\columnwidth]{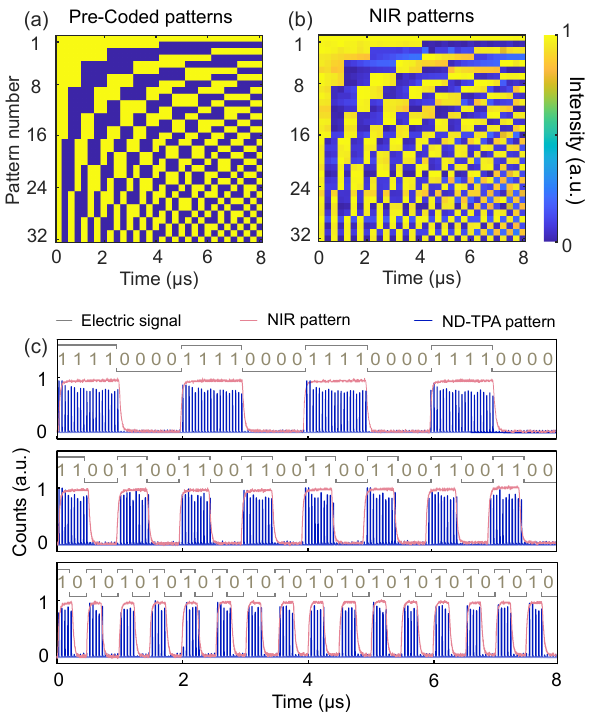}
	\caption{Mapping of temporal gating patterns to ND-TPA signals. (a) Encoding sequences for the 32-order Walsh-Hadamard matrix. (b) Measured matrix corresponding to the NIR modulation patterns. (c) ND-TPA responses for the 8th, 16th, and 32nd encoding patterns. The profiles in red represent the NIR encoding waveforms, while the envelopes for the burst of pulses in blue correspond to the mapped ND-TPA signals. The gray curves and binary digits indicate the bit sequences used to generate the temporal patterns.}
	\label{fig3}
\end{figure}

Beyond detector characterization, the fidelity of temporal encoding transfer through optical conversion and nonlinear detection is crucial for accurate TGI. Figure \ref{fig3}(a) presents the 32-order Walsh-Hadamard matrix used for temporal encoding, which was uploaded to the EOM operating at a 4 Mbps rate. The modulated NIR waveform after the OPA, shown in Fig. \ref{fig3}(b), was recorded by a 500 MHz InGaAs detector, confirming that the temporal modulation was well preserved during amplification. Figure \ref{fig3}(c) further compares representative encoding patterns (8th, 16th, and 32nd orders) with their corresponding ND-TPA responses. The red traces denote the NIR temporal modulation, while the blue profiles show the mapped ND-TPA signals measured by a high-speed Si detector (MenloSystems APD210, 400-1000 nm). The close temporal correspondence between the encoded waveforms and the ND-TPA envelopes verify that both the OPA and ND-TPA processes maintain high-fidelity temporal information transfer, ensuring consistent mapping of structured temporal patterns throughout the TGI system \cite{Wu2019Optica}.

\begin{figure*}[t!]
\includegraphics[width=0.83\textwidth]{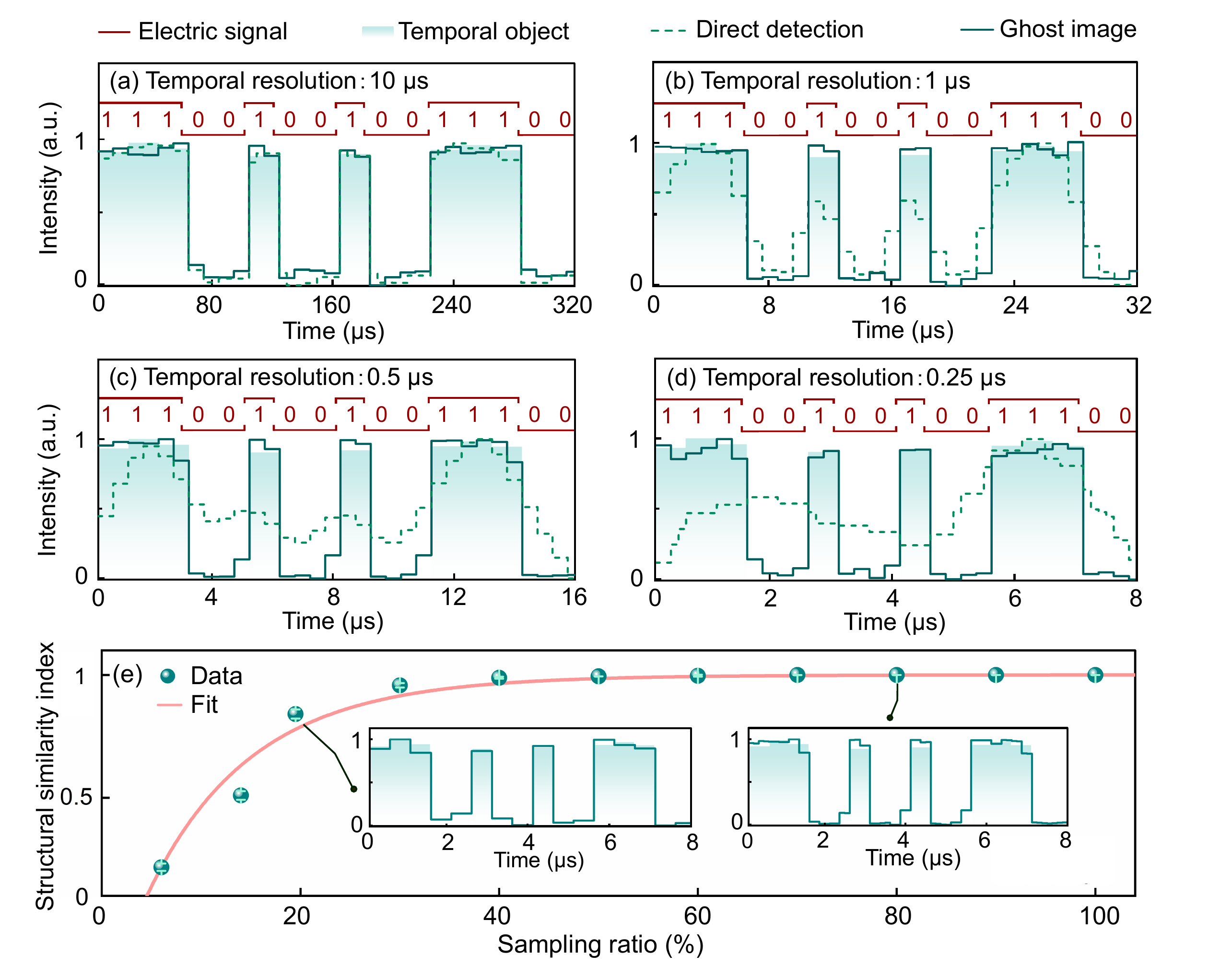}
	\caption{MIR TGI performances of binary temporal object and compressive sampling. (a-d) Reconstructed binary temporal sequences under preprogrammed encoding patterns at modulation rates of 0.1 Mbps (a), 1 Mbps (b), 2 Mbps (c), and 4 Mbps (d). The gray traces represent the electrical encoding signals applied to the MIR intensity modulator, the shaded regions indicate the ground-truth temporal object, the solid lines show the reconstructed waveforms, and the dashed lines correspond to direct detection by a 100 kHz silicon detector. The gray trace and binary digits represent the bit sequence used to generate the temporal object. The temporal resolution denotes the minimum resolvable temporal feature, determined by the pattern time-bin width. (e) Reconstruction fidelity, quantified by the structural similarity index (SSIM), as a function of the sampling ratio. The inset shows two representative results with compression ratios of 20\% and 80\%.}
	\label{fig4}
\end{figure*}

\subsection{MIR TGI via structured-pump ND-TPA}	
We next evaluated the performance of the MIR TGI system under structured optical pumping. Unlike conventional MIR TGI configurations that rely on nonlinear frequency mixing in external crystals, the ND-TPA approach offers a simpler and more cost-effective solution in which the photodetector simultaneously serves as both the nonlinear medium and the sensing element. As a result, the MIR temporal object is directly modulated and detected at the detector, without the need for phase matching, precise optical alignment, or temperature control. This intrinsic integration enables a compact, robust, and alignment-free system architecture \cite{Ma2025PhotoniX}.

The choice of encoding matrix plays a key role in determining the reconstruction fidelity. Compared with random modulation schemes \cite{Ryczkowski2016NP}, Walsh-Hadamard encoding provides orthogonal and balanced binary patterns that minimize crosstalk, distribute optical energy uniformly, and enhance tolerance to noise. These advantages improve reconstruction accuracy and accelerate data acquisition, thereby increasing overall imaging efficiency \cite{Devaux2016Optica}.

Figure \ref{fig4} summarizes the performance of binary TGI under different modulation rates and sampling conditions. The binary temporal object was generated by modulating the MIR pulses with an AOM and directly measured by a HgCdTe detector (UHSM-I-10.6, 3-12 $\mu$m) to represent the ground-truth waveform. As shown in Fig. \ref{fig4}(a), the MIR signal was modulated at 0.05 Mbps, while the structured NIR pump operates at 0.1 Mbps. For comparison, direct detection using a silicon photodiode (100 kHz bandwidth) was performed without encoding. The measured and reconstructed waveforms exhibit excellent agreement with the ground truth, validating the proposed TGI approach.

As the modulation rate increases, the finite bandwidth of the slow detector introduces strong distortion in direct detection; however, the TGI reconstruction remains accurate even beyond the detector's electronic bandwidth. As shown in Figs. \ref{fig4}(b-d), high-fidelity recovery of the temporal object is maintained at pump modulation rates of 1, 2, and 4 Mbps, demonstrating the robustness of ND-TPA-based TGI against bandwidth limitations. It should be noted that the temporal coding is implemented on an inter-pulse basis, so the ultimate encoding rate bound is set by the 20.1 MHz repetition rate. In this work, the MIR temporal object is defined by a MIR AOM with a maximum stable modulation speed of approximately 2 Mbps, so the maximum temporal resolution demonstrated is 4 Mbps, which could be increased toward the repetition rate limit with a higher speed MIR AOM.

Furthermore, compressive TGI is demonstrated in Fig. \ref{fig4}(e). This method enables accurate reconstruction with sampling ratios well below the Nyquist-Shannon limit. While traditional sampling requires at least twice the maximum signal frequency, compressive sensing allows faithful recovery from substantially fewer measurements \cite{Zhao2021Optica}. In the experiment, $M$ rows are selected from the $N \times N$ Walsh-Hadamard matrix to form the measurement matrix, yielding a sampling ratio of M/N. The temporal waveform is reconstructed using the Lagrangian TVAL3 algorithm, and its quality is evaluated by the structural similarity index (SSIM). As shown in Fig. \ref{fig4}(e), even at a 20\% sampling ratio, the SSIM remains above 80\%. Compared with conventional Nyquist-rate acquisition, the compressive approach reduces measurement time by more than fourfold while preserving high reconstruction fidelity.

\subsection{MIR photon-counting TGI}
We further investigated the performance of the MIR TGI system under photon-limited conditions, where a SPCM is employed to enhance detection sensitivity and ensure reliable signal acquisition. Unlike analog photodetectors that measure continuous photocurrents, the SPCM registers photon events as discrete electrical pulses, which are accumulated by a frequency counter over a defined integration window \cite{Zhang2025LPR}. In the experiment, photon counts corresponding to each temporal pattern were integrated within their respective time windows to yield a single intensity value per pattern, forming the measurement vector for correlation-based reconstruction.

\begin{figure}[b!]
\includegraphics[width=1\columnwidth]{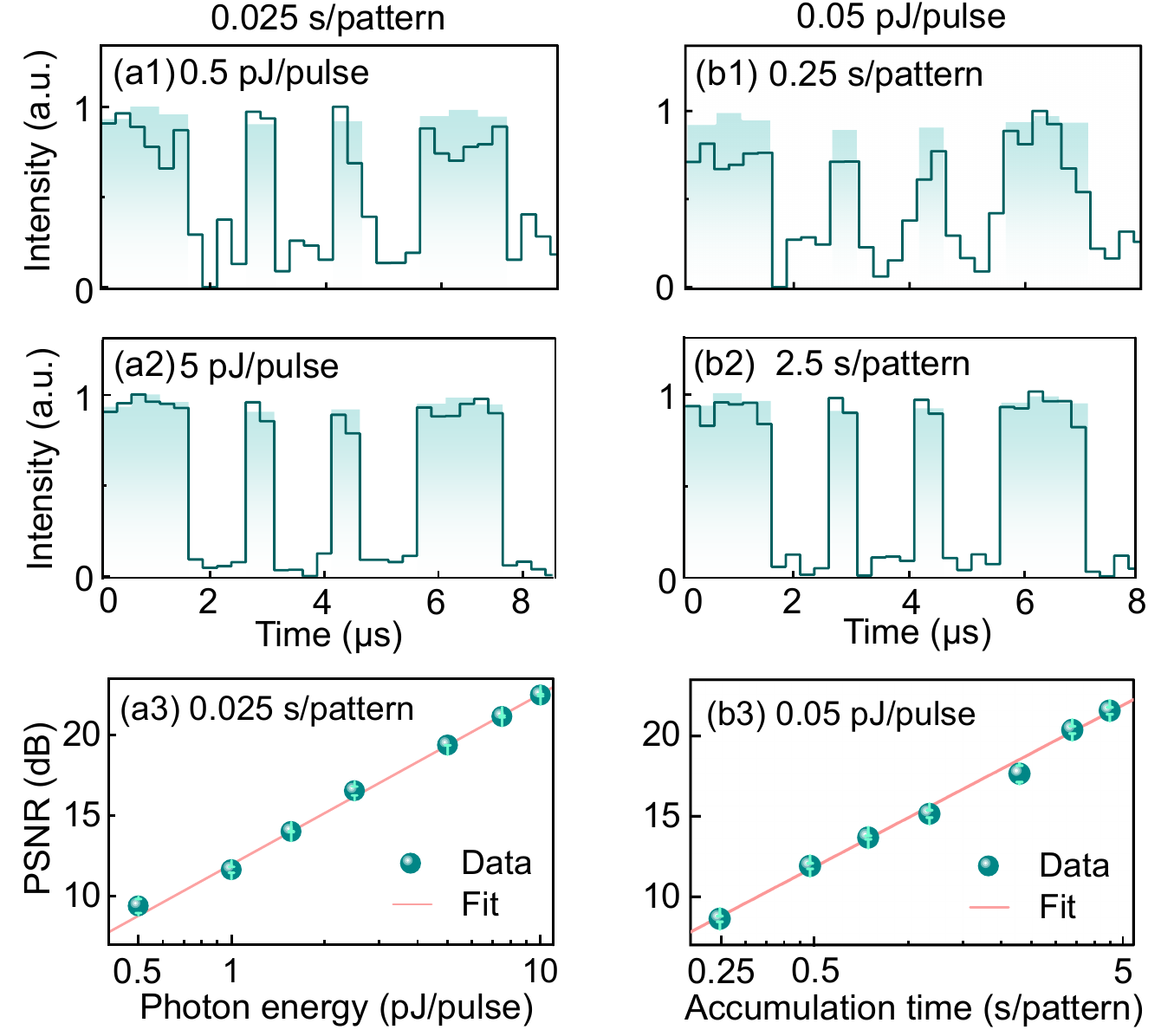}
	\caption{Characterization of single-photon MIR TGI performance. (a1-a2) Reconstructed temporal objects at a fixed integration time of 0.025 s per pattern with MIR pulse energies of 0.5 pJ (a1) and 5 pJ (a2), respectively. (a3) Peak signal-to-noise ratio (PSNR) as a function of the MIR pulse energy under a constant integration time of 0.025 s per pattern. (b1-b2) Reconstructed temporal objects at a fixed MIR pulse energy of 0.05 pJ with integration times of 0.25 s/pattern (b1) and 2.5 s/pattern (b2). (b3) PSNR as a function of the MIR pulse energy at a fixed integration time of 0.05 s/pattern.}
	\label{fig5}
\end{figure}
 
\begin{figure}[b!]
  \centering
  \includegraphics[width=0.8\columnwidth]{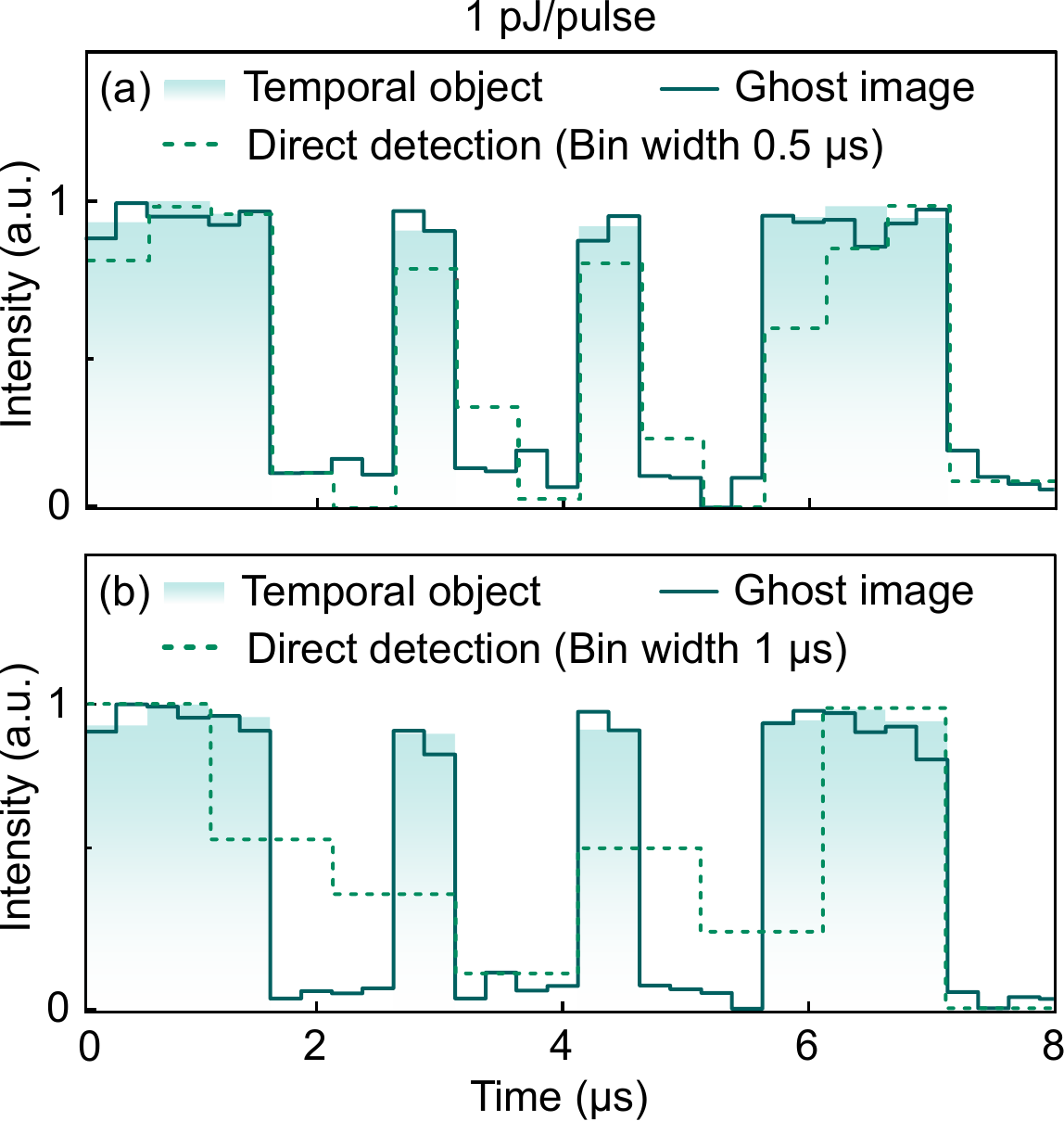}
  \caption{High-resolution reconstruction of single-photon MIR temporal waveforms. (a) Measurement results for a temporal object with a bit width of 0.5 $\mu$s (2 Mbps). The direct waveform measurement relies on the time-correlated single-photon counting technique. The binning width is set to be 0.5 $\mu$s for the histogram acquisition. In the ghost imaging measurement, the modulation resolution is 0.25 $\mu$s and the photon counts are integrated within a 8 $\mu$s time window. (b) The binning width in the direct detection scheme is set to be 1 $\mu$s, which is insufficient to accurately recover the temporal object.}
  \label{fig6}
\end{figure}

Figure \ref{fig5} presents the dependence of reconstruction fidelity on photon statistics. As shown in Figs. \ref{fig5}(a1-a2), with a fixed integration time of 0.025 s/pattern, increasing the MIR pulse energy improves the quality of the reconstructed waveform. At 0.5 pJ per pulse [Fig. \ref{fig5}(a1)], the reconstructed signal deviates from the ground truth, whereas at 5 pJ/pulse [Fig. \ref{fig5}(a2)], the recovered temporal steps closely match the reference. The quantitative dependence, evaluated using the peak signal-to-noise ratio (PSNR) \cite{Xu2018OE, Wu2024LSA, Zhang2025LPR}, exhibits a monotonic rise with pulse energy [Fig. \ref{fig5}(a3)], reflecting the photon-counting nature of the process, $\textit{i.e.,}$ higher photon flux suppresses shot noise and stabilizes reconstruction.

A similar trend is observed for varying integration times. As shown in Figs. \ref{fig5}(b1-b2), at a fixed MIR pulse energy of 0.05 pJ, extending the integration time from 0.25 s to 2.5 s per pattern leads to smoother and more accurate reconstructions. The corresponding PSNR values [Fig. \ref{fig5}(b3)] increase proportionally, confirming that reconstruction fidelity is governed by the effective photon budget. These results reveal an inherent trade-off between reconstruction quality and acquisition speed in photon-limited ghost imaging, which may be mitigated through faster modulation or advanced reconstruction algorithms \cite{Xu2018OE, Zhao2021Optica}. The achieved MIR detection sensitivity underscores the advantage of ND-TPA-enabled photon counting in the MIR.

We further demonstrated that single-photon MIR TGI could surpass the effective timing-resolution limit of conventional photon-counting detection. To examine this, a time-correlated single-photon counting (TCSPC) module is combined with the SPCM to emulate different temporal resolving windows by adjusting the bin width. As shown in Fig. \ref{fig6}(a), at a bin width of 0.5 $\mu$s (matching the 2-MHz modulation speed of the temporal waveform), both direct detection and TGI accurately recover the temporal waveform. In contrast to conventional time-resolved photon detection, which relies on precise event timing, the TGI method retrieves temporal information statistically through correlations between photon counts and encoded patterns. Each encoded pattern is integrated over an 8-$\mu$s window with a modulation resolution of 0.25 $\mu$s, converting time-resolved variations into statistically correlated intensity measurements.

When the bin width was relaxed to 1 $\mu$s [Fig. \ref{fig6}(b)], direct detection became severely distorted due to undersampling, whereas TGI still reconstructed the waveform with high fidelity. These results confirm that ND-TPA-based TGI overcame the temporal bandwidth constraints inherent to photon-counting detection by encoding temporal information into statistical correlations rather than relying on electronic timing precision. Notably, the use of TCSPC in this demonstration served only as a diagnostic tool. Practically, the entire experiment can be implemented using a standard frequency counter, highlighting the simplicity, scalability, and low-cost potential of this approach for broadband MIR TGI.

\subsection{Broadband performance of MIR ND-TPA TGI}
\begin{figure}[t!]
\includegraphics[width=0.95\columnwidth]{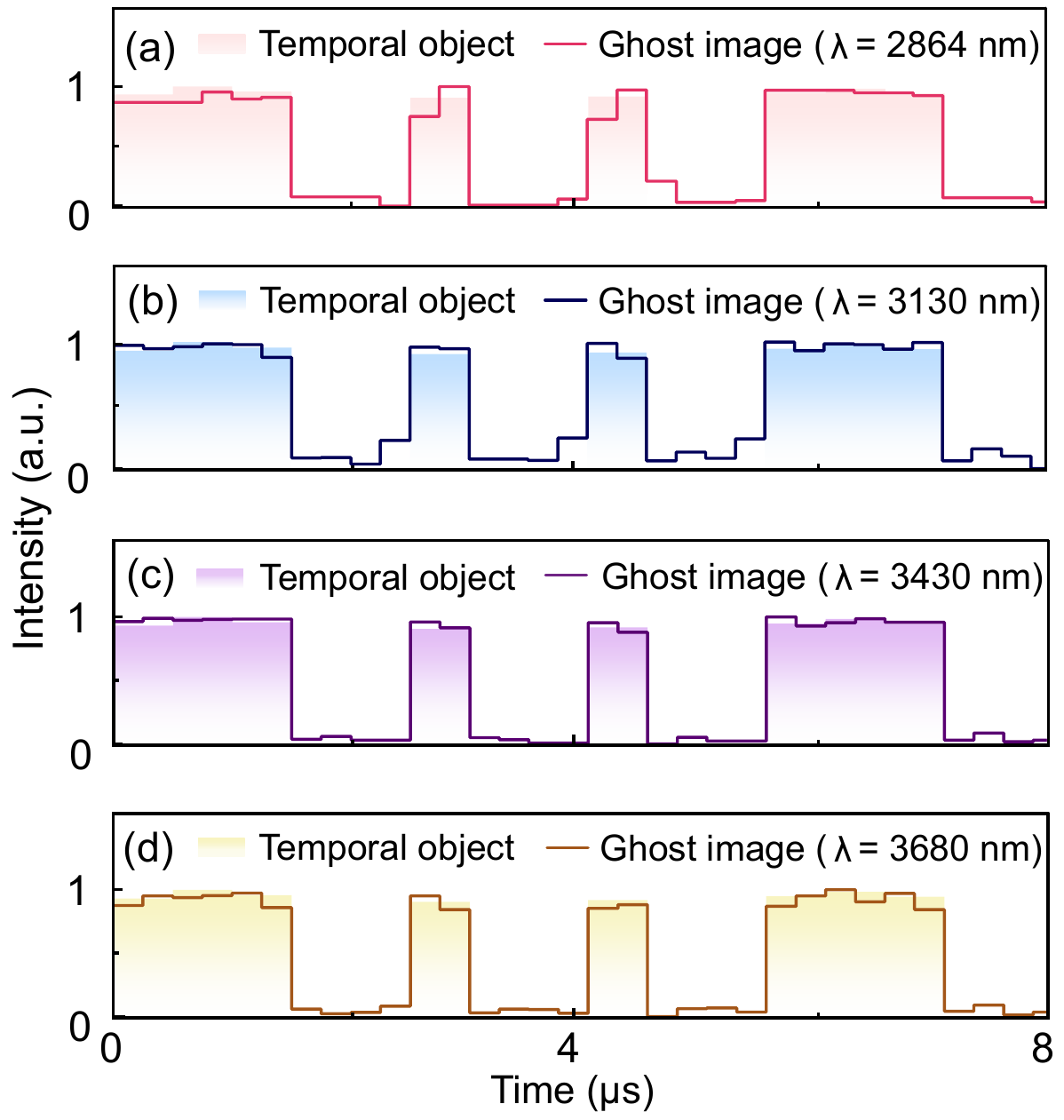}
	\caption{Reconstructed temporal waveforms obtained by tuning the MIR signal wavelength at 2864 nm (a), 3130 nm (b), 3430 nm (c), and 3680 nm (d). The consistent reconstruction across a wide spectral range demonstrates the broadband capability of the ND-TPA-based TGI system.}
	\label{fig7}
\end{figure}

Finally, we extended the MIR ND-TPA TGI system to multispectral operation, highlighting its inherent broadband capability. This broadband response originates from the ND-TPA process occurring directly at the detector surface, which removes the need for phase matching and naturally supports wide spectral coverage \cite{Cirloganu2011OE}. Figure \ref{fig7} presents reconstructed temporal signals obtained by tuning the MIR signal wavelength while keeping the NIR pump fixed. High-fidelity reconstructions are achieved at representative wavelengths of 2864 nm, 3130 nm, 3430 nm, and 3680 nm, demonstrating consistent performance across a broad spectral range. The results confirm that the system maintains stable operation without optical re-optimization when the signal wavelength is varied. The tuning range is primarily bounded by the ND-TPA energy-sum condition in silicon under a fixed 1550 nm pump, the combined photon energy must exceed the bandgap, which sets a practical upper limit on the MIR wavelength. Meanwhile, the pump wavelength should lie outside the detector’s one-photon absorption band to avoid direct pump absorption dominating the response.

These findings verify that ND-TPA endows MIR TGI with intrinsic spectral flexibility, enabling reliable multispectral and broadband temporal imaging without modifying the optical configuration. This characteristic positions ND-TPA TGI as a promising approach for broadband MIR spectroscopy, hyperspectral sensing, and ultrafast molecular detection. The spectral reach can be extended to longer MIR wavelengths in future by using a shorter-wavelength pump or a detector material with a more suitable bandgap \cite{Ma2025PhotoniX}.

\section{Conclusion}        
In conclusion, we have demonstrated a broadband MIR TGI system based on temporally structured pumped ND-TPA. Through the interaction between the MIR signal and the NIR structured pump at a large bandgap detector, this approach effectively extends the spectral response of high performance silicon photodiodes into the MIR regime. Compared with crystal-based MIR TGI schemes \cite{Wu2024LSA,Zhang2025LPR}, it eliminates the need for phase matching, inherently supports broadband operation, and significantly simplifies the optical configuration. The reconstructed temporal profiles exhibit high contrast and low noise, while the detection remains efficient even under weak MIR illumination fluxes as small as 0.05 pJ/pulse. Moreover, robust temporal reconstruction is maintained across a wide range of MIR wavelengths without the need for optical re-optimization. This spectral flexibility underscores the inherent broadband nature of the ND-TPA process and its suitability for compact, broadband MIR TGI architectures. Collectively, these results verify the experimental feasibility of MIR TGI via ND-TPA and highlight its potential for broadband, high-sensitivity MIR detection and imaging applications.

Future improvements can be pursued along several directions. The attainable temporal reconstruction resolution is primarily determined by the bandwidth of the employed NIR modulators. Increasing modulation bandwidths will enable shorter pattern durations, more accurate temporal gating, and reduced inter-slot crosstalk. Nowadays, commercial EOMs allow high operation bandwidths beyond 40 GHz, making it feasible to resolve MIR temporal objects at rates up to tens of Gbps \cite{Wu2024LSA}. In parallel, optimizing the pump wavelength and/or replacing silicon with direct-bandgap semiconductors such as InGaAs \cite{Potma2021APLP} or GaN \cite{Fishman2011NP} could extend the operation spectral range, increase ND-TPA absorption efficiency and improve detection sensitivity. In addition, adaptive encoding and compressed sensing based reconstruction \cite{Zhao2021Optica} can further enhance acquisition efficiency and noise robustness. These developments will enable the proposed MIR TGI paradigm to stably and accurately reconstruct transient temporal signals, maintain robustness under photon-starved or noisy conditions, and expand its use in broadband MIR sensing and communication.

\section*{Acknowledgements}
This work was funded by Shanghai Pilot Program for Basic Research (TQ20220104); National Natural Science Foundation of China (62175064, 62235019); Postdoctoral Fellowship Program (GZC20250545); China Postdoctoral Science Foundation (2024M760918, 2025T180224).
	
\section*{Conflict of Interest}
The authors declare no conflict of interests.
	
\section*{Data Availability Statement}
The data that support the findings of this study are available from the corresponding author upon reasonable request.
	
\section*{Keywords}
temporal ghost imaging, mid-infrared detection, two-photon absorption, nonlinear optical modulation

\end{document}